\documentclass[
    twocolumn,
	prd,
	amssymb,
	preprintnumbers,superscriptaddress,
	nofootinbib]{revtex4-1}

\pdfoutput=1
\usepackage{paralist}
\usepackage{graphicx}
\usepackage{enumitem}
\usepackage{latexsym}
\usepackage{amsfonts}
\usepackage{amssymb}
\usepackage{xcolor}
\usepackage[export]{adjustbox}
\usepackage{amsmath}
\usepackage[thinlines]{easytable}
\usepackage{slashed}
\usepackage{dcolumn}
\usepackage{verbatim}
\usepackage{float}
\usepackage{multirow}
\usepackage{xspace}
\usepackage[normalem]{ulem}
\usepackage[
pdfauthor={Jeremy Sakstein}]{hyperref}
\usepackage{tabularx}
\usepackage{lettrine}
\usepackage{setspace}

\input Zallman.fd

\LettrineTextFont{\itshape}

\newcommand{\beq}{\begin{equation}}
\newcommand{\eeq}{\end{equation}}
\newcommand{\bea}{\begin{eqnarray}}
\newcommand{\eea}{\end{eqnarray}}

\newcommand{\alg}[1]{\begin{align} \begin{split} #1 \end{split}  \end{align}}

\newcommand{\refjnl}[1]{{\rm#1}}

\def\aj{\refjnl{AJ}}                   
\def\apj{\refjnl{ApJ}}                 
\def\aap{\refjnl{A\&A}}                

\def\mnras{\refjnl{MNRAS}}             
\usepackage{pdfbase}[2017/03/16]
\usepackage{xparse,ocgbase}
\usepackage{xcolor,calc}
\usepackage{tikzpagenodes,linegoal}
\usetikzlibrary{calc}
\usepackage{tcolorbox}

\ExplSyntaxOn
\let\tpPdfLink\pbs_pdflink:nn
\let\tpPdfAnnot\pbs_pdfannot:nnnn\let\tpPdfLastAnn\pbs_pdflastann:
\let\tpAppendToFields\pbs_appendtofields:n
\def\tpPdfXform{\pbs_pdfxform:nnnnn{1}{1}{}{}}
\let\tpPdfLastXform\pbs_pdflastxform:
\let\cListSet\clist_set:Nn\let\cListItem\clist_item:Nn
\ExplSyntaxOff

\usepackage{pdfbase}[2017/03/16]
\usepackage{xparse,ocgbase}
\usepackage{xcolor,calc}
\usepackage{tikzpagenodes,linegoal}
\usetikzlibrary{calc}
\usepackage{tcolorbox}

\ExplSyntaxOn
\let\tpPdfLink\pbs_pdflink:nn
\let\tpPdfAnnot\pbs_pdfannot:nnnn\let\tpPdfLastAnn\pbs_pdflastann:
\let\tpAppendToFields\pbs_appendtofields:n
\def\tpPdfXform{\pbs_pdfxform:nnnnn{1}{1}{}{}}
\let\tpPdfLastXform\pbs_pdflastxform:
\let\cListSet\clist_set:Nn\let\cListItem\clist_item:Nn
\ExplSyntaxOff

\makeatletter
\NewDocumentCommand{\tooltip}{%
  ssssO{\ifdefined\@linkcolor\@linkcolor\else blue\fi}mO{yellow!20}mO{0pt,0pt}%
}{{%
  \leavevmode%
  \IfBooleanT{#2}{%
    \ocgbase@new@ocg{tipOCG.\thetcnt}{%
      /Print<</PrintState/OFF>>/Export<</ExportState/OFF>>%
    }{false}%
    \xdef\tpTipOcg{\ocgbase@last@ocg}%
    \ocgbase@add@ocg@to@radiobtn@grp{tool@tips}{\ocgbase@last@ocg}%
  }%
  \tpPdfLink{%
    \IfBooleanTF{#4}{%
      /Subtype/Link/Border[0 0 0]/A <</S/SetOCGState/State [/Toggle \tpTipOcg]>>
    }{%
      /Subtype/Screen%
      /AA<<%
        \IfBooleanTF{#3}{%
          /E<</S/SetOCGState/State [/Toggle \tpTipOcg]>>%
        }{%
          \IfBooleanTF{#2}{%
            /E<</S/SetOCGState/State [/ON \tpTipOcg]>>%
            /X<</S/SetOCGState/State [/OFF \tpTipOcg]>>%
          }{
            \IfBooleanTF{#1}{%
              /E<</S/JavaScript/JS(%
                var fd=this.getField('tip.\thetcnt');%
                if(typeof(click\thetcnt)=='undefined'){%
                  var click\thetcnt=false;%
                  var fdor\thetcnt=fd.rect;var dragging\thetcnt=false;%
                }%
                if(fd.display==display.hidden){%
                  fd.delay=true;fd.display=display.visible;fd.delay=false;%
                }else{%
                  if(!click\thetcnt&&!dragging\thetcnt){fd.display=display.hidden;}%
                  if(!dragging\thetcnt){click\thetcnt=false;}%
                }%
                this.dirty=false;%
              )>>%
            }{%
              /E<</S/JavaScript/JS(%
                var fd=this.getField('tip.\thetcnt');%
                if(typeof(click\thetcnt)=='undefined'){%
                  var click\thetcnt=false;%
                  var fdor\thetcnt=fd.rect;var dragging\thetcnt=false;%
                }%
                if(fd.display==display.hidden){%
                  fd.delay=true;fd.display=display.visible;fd.delay=false;%
                }%
               this.dirty=false;%
              )>>%
              /X<</S/JavaScript/JS(%
                if(!click\thetcnt&&!dragging\thetcnt){fd.display=display.hidden;}%
                if(!dragging\thetcnt){click\thetcnt=false;}%
                this.dirty=false;%
              )>>%
            }%
            /U<</S/JavaScript/JS(click\thetcnt=true;this.dirty=false;)>>%
            /PC<</S/JavaScript/JS (%
              var fd=this.getField('tip.\thetcnt');%
              try{fd.rect=fdor\thetcnt;}catch(e){}%
              fd.display=display.hidden;this.dirty=false;%
            )>>%
            /PO<</S/JavaScript/JS(this.dirty=false;)>>%
          }%
        }%
      >>%
    }%
  }{{\color{#5}#6}}%
  \sbox\tiptext{%
    \IfBooleanT{#2}{%
      \ocgbase@oc@bdc{\tpTipOcg}\ocgbase@open@stack@push{\tpTipOcg}}%
    \tcbox[colframe=black,colback=#7,size=fbox,arc=1ex,sharp corners=southwest]{#8}%
    \IfBooleanT{#2}{\ocgbase@oc@emc\ocgbase@open@stack@pop\tpNull}%
  }%
  \cListSet\tpOffsets{#9}%
  \edef\twd{\the\wd\tiptext}%
  \edef\tht{\the\ht\tiptext}%
  \edef\tdp{\the\dp\tiptext}%
  \tipshift=0pt%
  \IfBooleanTF{#2}{%
    \setlength\whatsleft{\linegoal}%
  }{%
    \measureremainder{\whatsleft}%
  }%
  \ifdim\whatsleft<\dimexpr\twd+\cListItem\tpOffsets{1}\relax%
    \setlength\tipshift{\whatsleft-\twd-\cListItem\tpOffsets{1}}\fi%
  \IfBooleanF{#2}{\tpPdfXform{\tiptext}}%
  \raisebox{\heightof{#6}+\tdp+\cListItem\tpOffsets{2}}[0pt][0pt]{%
    \makebox[0pt][l]{\hspace{\dimexpr\tipshift+\cListItem\tpOffsets{1}\relax}%
    \IfBooleanTF{#2}{\usebox{\tiptext}}{%
      \tpPdfAnnot{\twd}{\tht}{\tdp}{%
        /Subtype/Widget/FT/Btn/T (tip.\thetcnt)%
        /AP<</N \tpPdfLastXform>>%
        /MK<</TP 1/I \tpPdfLastXform/IF<</S/A/FB true/A [0.0 0.0]>>>>%
        /Ff 65536/F 3%
        /AA <<%
          /U <<%
            /S/JavaScript/JS(%
              var fd=event.target;%
              var mX=this.mouseX;var mY=this.mouseY;%
              var drag=function(){%
                var nX=this.mouseX;var nY=this.mouseY;%
                var dX=nX-mX;var dY=nY-mY;%
                var fdr=fd.rect;%
                fdr[0]+=dX;fdr[1]+=dY;fdr[2]+=dX;fdr[3]+=dY;%
                fd.rect=fdr;mX=nX;mY=nY;%
              };%
              if(!dragging\thetcnt){%
                dragging\thetcnt=true;Int=app.setInterval("drag()",1);%
              }%
              else{app.clearInterval(Int);dragging\thetcnt=false;}%
              this.dirty=false;%
            )%
          >>%
        >>%
      }%
      \tpAppendToFields{\tpPdfLastAnn}%
    }%
  }}%
  \stepcounter{tcnt}%
}}
\makeatother
\newsavebox\tiptext\newcounter{tcnt}
\newlength{\whatsleft}\newlength{\tipshift}
\newcommand{\measureremainder}[1]{%
  \begin{tikzpicture}[overlay,remember picture]
    \path let \p0 = (0,0), \p1 = (current page.east) in
      [/utils/exec={\pgfmathsetlength#1{\x1-\x0}\global#1=#1}];
  \end{tikzpicture}%
}

\newcommand{\msun}{{\rm M}_\odot}

\DeclareRobustCommand{\okina}{%
  \raisebox{\dimexpr\fontcharht\font`A-\height}{%
    \scalebox{0.8}{`}%
  }%
}

\allowdisplaybreaks

\begin{document}

\title{Axions and the RR Lyrae Period-Luminosity Relation}

\author{Jeremy Sakstein} \email{sakstein@hawaii.edu}
\affiliation{Department of Physics \& Astronomy, University of Hawai\okina i, Watanabe Hall, 2505 Correa Road, Honolulu, HI, 96822, USA}

\author{Djuna Croon} \email{djuna.l.croon@durham.ac.uk}
\affiliation{Institute for Particle Physics Phenomenology, Department of Physics, Durham University, Durham DH1 3LE, U.K.}

\date{\today}

\begin{abstract}
We investigate the impact of axions on the period--luminosity relation (PLR) of RR Lyrae stars. During the red giant branch phase, emission through the electron coupling cools the core, increasing its mass at ignition. The larger core produces more luminous horizontal-branch stars, systematically shifting the zero point of the RR Lyrae PLR toward brighter magnitudes. We model the RR~Lyrae population of the globular cluster M3 using stellar evolution and pulsation calculations tailored to its age and composition, and compare the resulting synthetic $K_s$-band PLR with observations.~For $\alpha_{26}\equiv 10^{26}g_{ae}^2/4\pi=0.5$, the axion-induced shift in the predicted PLR is comparable to the current uncertainty in the absolute M3 zero-point calibration.
\end{abstract}

\preprint{IPPP/26/66}

\maketitle

\section{Introduction}

Axions, or axion-like particles (ALPs) more generally, are light pseudoscalar particles that are motivated by a variety of beyond the Standard Model (BSM) scenarios e.g., the QCD axion  \cite{Peccei:1977hh,Weinberg:1977ma,Wilczek:1977pj,Zhitnitsky:1980tq}, the pions of a dark standard model \cite{Maleknejad:2022gyf, Alexander:2023wgk}, and the axion-like particles predicted by string theory (see e.g., \cite{Svrcek:2006yi, Arvanitaki:2009fg}).~They naturally couple to Standard Model (SM) particles such as electrons, photons, and nucleons, leading to a plethora of observational signatures that have been searched for experimentally (see e.g., the reviews \cite{Irastorza:2018dyq,DiLuzio:2020wdo}).

Light axions ($m_a\lesssim10$ keV) can be produced copiously in stellar interiors.~In the weakly coupled regime they subsequently free-stream out of the star, acting as a source of energy loss analogous to neutrino emission.~This has observable consequences across various stellar types, which have been used to exclude regions of the axion-SM coupling parameter space (see \cite{Caputo:2024oqc,Carenza:2024ehj} for recent reviews).

Observations of stars at the tip of the red giant branch (TRGB) provide one of the leading probes of the axion--electron coupling \cite{Viaux:2013lha,Capozzi:2020cbu,Straniero:2020iyi,Dennis:2023aam,Troitsky:2024keu,Gontcharov:2026knx}.~During the red giant branch phase, axion emission cools the core, delaying the helium flash.~This leaves more time for the hydrogen-burning shell to deposit fresh helium onto the core, increasing its mass and thereby brightening the tip of the red giant branch \cite{Raffelt:1989xu,Raffelt:1990yz,Raffelt:1994ry,Raffelt:1996wa}.~Comparing stellar-evolution calculations with empirical calibrations of the TRGB I-band magnitude therefore constrains the coupling, although the inference depends on the treatment of stellar and observational systematics.~Published limits probe $\alpha_{26}\sim\mathcal{O}(0.01\text{--}1)$, depending on the adopted stellar modeling, calibration strategy, and treatment of systematics \cite{Capozzi:2020cbu,Straniero:2020iyi,Dennis:2023aam,Gontcharov:2026knx}.~The most recent analysis, Ref.~\cite{Gontcharov:2026knx}, reports $\alpha_{26}\lesssim0.01$ using bolometric TRGB luminosities rather than the $I$-band magnitude, thereby avoiding the dependence on the theoretical effective temperature and bolometric correction that contributes to the uncertainty in $M_I$-based analyses.

Here, we investigate whether the increased core mass can lead to additional observational signatures from post-flash objects.~After the flash, the star settles onto the horizontal branch (HB), where it burns helium in its core.~Depending on their masses and envelope properties, some of these objects become RR~Lyrae stars, exhibiting large-amplitude pulsations with a tight period--luminosity relation (PLR).~The luminosities and pulsation periods of RR~Lyrae variables have previously been used to constrain the helium-core mass and exotic cooling \cite{Raffelt:1994ry,Catelan:1995ba}.~Here, we investigate the PLR as a direct observable of the axion-induced increase in the helium-core mass.~Using stellar modeling software, we simulate the evolution of RR~Lyrae stars under axion--electron losses, finding that the synthetic PLR is systematically shifted to brighter magnitudes.~Applying this framework to the globular cluster M3, where the PLR has recently been calibrated using a sample of 233 RR~Lyrae stars \cite{2020AJ....160..220B}, we estimate sensitivity to couplings of order $\alpha_{26}\sim0.5$.~Our results therefore indicate that RR~Lyrae PLRs may provide a complementary post-helium-flash probe of axion-induced RGB cooling, although their sensitivity to the axion--electron coupling is weaker than the strongest current TRGB constraints.

The remainder of this paper is organized as follows.~We review the formation and properties of RR Lyrae stars in section~\ref{sec:RRLyrae}.~Section~\ref{sec:ax-emission} reviews axion-like particles and their coupling to electrons, and details the emission channels and rates used in this work.~Our procedure for simulating RR Lyrae stars and predicting their period--luminosity relation is presented in section~\ref{sec:stellar_response}.~We present our results in section~\ref{sec:results} before concluding in section~\ref{sec:conclusions}.

\section{RR Lyrae Stars}
\label{sec:RRLyrae}

RR~Lyrae stars are low-mass ($\sim$$0.6$--$0.8\,\msun$), core-helium-burning stars that occupy the instability strip region of the Hertzsprung–Russell (HR) diagram.~They are found primarily in old stellar populations such as globular clusters and the stellar halo, with ages $\gtrsim 10$~Gyr.

RR~Lyrae stars descend from low-mass ($M\lesssim2\msun$) stars that have evolved through the red giant branch (RGB). During this phase, hydrogen-shell burning grows an inert helium core, while stellar winds remove part of the hydrogen-rich envelope. Once the core reaches a critical mass, helium ignites under degenerate conditions in a \textit{helium flash}. The star subsequently settles onto the horizontal branch, where it burns helium in its core.~The luminosity of an HB star is determined primarily by its helium-core mass, while its effective temperature depends largely on the mass of the hydrogen envelope.  

Stars with suitable envelope masses occupy the \textit{instability strip} and become RR~Lyrae variables. The instability strip is a near-vertical band in the Hertzsprung--Russell diagram where radial pulsations are self-excited by the $\kappa$-mechanism.~RR~Lyrae envelopes contain a partial helium-ionization zone where $\mathrm{He}^{+}$ is ionized to $\mathrm{He}^{++}$.~Stellar compression increases the ionization fraction and opacity, temporarily trapping radiation within the star. The resulting pressure buildup drives the layer outward, after which cooling and recombination reduce the opacity and allow radiation to escape. The layer subsequently contracts, restarting the cycle and sustaining coherent pulsations.~

Both the fundamental and first-overtone modes are observed.~Stars pulsating in the former are classified as RRab stars, with periods typically ranging from $0.4$ to $0.8$~days, while those pulsating in the latter are classified as RRc stars, with shorter periods of $\sim 0.2$--$0.5$~days.~Some objects, known as RRd stars, pulsate in both modes simultaneously.~

\section{Axions in Stellar Interiors}
\label{sec:ax-emission}

In this section we review the relevant features of axions coupled to electrons and their production in stars.~We refer the reader to e.g., \cite{Croon:2020oga,Croon:2022rrv}, and references therein for more details.

The Lagrangian for an axion-like particle $a$ with mass $m_a$ coupled to electrons is
\begin{align} \label{Lae}
    \mathcal{L} = \mathcal{L}_{\rm SM}+\frac12\partial_\mu a\partial^\mu a - i g_{ae} a \bar\psi_e \gamma_5 \psi_e  - \frac12 m_a^2 a^2,
\end{align}
where $\mathcal{L}_{\rm SM}$ is the SM Lagrangian,  $g_{ae}$ is the axion-electron Yukawa coupling, and $\psi_e$ are the electron Dirac spinors.~We neglect the mass henceforth since we are interested in ALPs with masses lighter than the temperatures of RR Lyrae stars and their progenitors.~By convention, we will work with the quantity $\alpha_{26} \equiv 10^{26} \alpha_{ae} \equiv 10^{26} g^2_{ae}/4\pi$, which we refer to as the axion-electron coupling.

Axions in RGB stars are primarily produced via semi-Compton and bremsstrahlung processes.~The energy loss rate due to semi-Compton scattering, $e+\gamma \rightarrow e+a$, is given by \cite{Raffelt:1994ry},
\begin{eqnarray}\label{eq:Q_SC}
    \mathcal Q_{\rm sC} \! = \! \frac{160 \,\zeta_6 \alpha_{\rm EM} \alpha_{ae}}{\pi}  \, \frac{Y_e T^6F_{\rm deg}}{m_N m_e^4} \!\simeq
    33\alpha_{26}Y_eT_8^6F_{\rm deg} {\rm \frac{erg}{g \!\cdot\! s} },~~~~
\end{eqnarray}
where $\zeta_6 = \pi^6/945$, $\alpha_{\rm EM} = 1/137$ is the electromagnetic fine-structure constant, $Y_e = Z/A $ is the number of electrons per baryon, $m_N,m_e$ are the nucleon and electron mass respectively, and $T_8=(T/10^8\textrm{K})$. The function $F_{\rm deg}$ encodes the Pauli-blocking of the process due to electron degeneracy.~A good numerical approximation for $F_{\rm deg}$ is \cite{Croon:2020oga}:
\alg{
    F_{\rm deg} &= \frac{1}{2} \left[ 1-\tanh  f(\rho,T) \right] \\ f(\rho,T) &=
    a \log_{10} \left[ \frac{\rho}{\text{g cm}^{-3}}\right] -b \log_{10}\left[ \frac{T}{\rm K}\right] +c,
}
with $a=0.973$, $b=1.596$, and $c=8.095$.

The energy loss rate due to bremsstrahlung  $e+ (Z,A) \to e + (Z,A) + a$ depends on the electron degeneracy.~For non-relativistic electrons, the ALP bremsstrahlung rate in the non-degenerate (ND) and degenerate (D) regimes are \cite{Raffelt:1994ry}
\begin{eqnarray}
    \label{eq:Q_bremm}
    &&\mathcal Q_{b,{\rm ND}} \!=\! \frac{128}{45} \frac{\alpha_{\rm EM}^2 \alpha_{ae} \rho T^{5/2}}{\sqrt{\frac\pi2} m_N^2 m_e^{7/2}} F_{b,{\rm ND}} \! \simeq \! 0.58 \alpha_{26} {\rm \frac{erg}{g \! \cdot \! s} } \rho_3 T_8^{5/2} F_{b,{\rm ND}} \nonumber
    \\ 
    &&\mathcal Q_{b,{\rm D}} \!=\!\frac{\pi^2}{15}\frac{Z^2}{A}\frac{\alpha_{\rm EM}^2 \alpha_{ae} T^4}{m_N m_e^2} F_{b,{\rm D}} \simeq 10.8\, \alpha_{26} {\rm \frac{erg}{g \! \cdot \! s} }  T_8^4 F_{b,{\rm D}}, 
\end{eqnarray}
where $\rho_3 = \rho/(10^3{\rm g/cm^3})$, $F_{b,{\rm ND}}  = Z(1+Z)/A$ and
\begin{align}
F_{b,{\rm D}} &\nonumber= \frac{2}{3} \log\left(\frac{2+\kappa^2}{\kappa^2} \right)\\& + \left[ \left(\kappa^2 +\frac25 \right) \log\left( \frac{2+\kappa^2}{\kappa^2} \right) - 2\right] \frac{\beta_F^2}3.\label{eq:Fbd}
\end{align}
In this expression, $\beta_F=p_F/E_F$ is the electron Fermi velocity and
\begin{align}
\kappa^2 \equiv \frac{k_S^2}{2p_F^2}
\end{align}
is a dimensionless screening parameter where  $k_S$ is the Debye screening momentum,
\begin{align}
k_S^2 = \frac{4\pi\alpha_{\rm EM}}{T}\sum_i n_i Z_i^2,
\label{debye-momentum}
\end{align}
where $i$ runs over all ion species in the plasma.

Combining these processes, the total specific energy loss rate from electrophilic axion emission is \cite{Raffelt:1994ry}
\beq
\mathcal Q_{ae} = \mathcal Q_{\rm sC} + ( \mathcal Q_{b,{\rm ND}}^{-1} + \mathcal Q_{b,{\rm D}}^{-1})^{-1}.
\eeq

\section{Modeling RR~Lyrae Stars Under Axion Cooling}
\label{sec:stellar_response}

In this section we describe our methodology for simulating the RR Lyrae period-luminosity relation.

\subsection{Theoretical Considerations}

Before proceeding to detailed numerical modeling, it is instructive to first gain some theoretical insight into the RR Lyrae PLR. 

The PLR arises as follows.~The period of stellar modes scale as $P\sim \sqrt{R^3/GM}$ up to an $\mathcal{O}(1)$ coefficient \cite{1968pss..book.....C,2004cgps.book.....W}, so that 
\begin{equation}
\log P = \frac{3}{2}\log R - \frac{1}{2}\log M+\alpha,
\end{equation}
where $\alpha$ is a mode-dependent constant.~Using the Stefan--Boltzmann relation, $L=4\pi R^2\sigma T_{\rm eff}^4$, this may be written as
\begin{equation}
\log P =
\frac{3}{4}\log L
-3\log T_{\rm eff}
-\frac{1}{2}\log M
+\beta,
\label{eq:PLR_theory1}
\end{equation}
where $\beta$ is a second constant.~

In practice, observations are made in some photometric band X, so the luminosity in equation \eqref{eq:PLR_theory1} must be converted to the absolute magnitude in this band $M_X$
\begin{equation}
M_X=M_{\rm bol}-BC_X(T_{\rm eff},\log g,[{\rm Fe/H}]),
\end{equation}
where $BC_X$ is the corresponding bolometric correction (BC).~Bands where the $T_{\rm eff}$-dependence of the BC compensate for that of the dependence in \eqref{eq:PLR_theory1} produce tight PLR relations, with some scatter due to the metallicity-dependence.~The near infra-red is one example.~In this work we focus on the $K_s$ band, where M3 has homogeneous time-series RR~Lyrae photometry and a well-defined empirical PLR \cite{2020AJ....160..220B}.

Equation~\eqref{eq:PLR_theory1} helps to guide our expectations for how axions affect the RR Lyrae PLR.~The luminosity of an HB star is primarily controlled by its helium-core mass and, as noted above, axion cooling during RGB evolution increases this. We therefore expect axions to induce a systematic shift in the RR~Lyrae PLR toward brighter magnitudes at fixed period.~This would manifest as a shift in the PLR zero-point.

In what follows, we will focus exclusively on the fundamental mode. This gives the most direct comparison between the computed pulsation periods and the observed RRab PLR, and avoids introducing an additional fundamentalization prescription for first-overtone RRc stars. 

\subsection{Modeling RR Lyrae Stars}

RR Lyraes inhabit a diverse range of environments.~Given the theoretical considerations above, we expect a system-to-system variation in the PLR due to age and metallicity, so it is important that the stellar modeling choices are informed by the system under consideration.~Therefore, we first select a host system to study and choose stellar modeling parameters compatible with this.

\subsubsection{RR Lyrae Host System}

We chose the globular cluster M3 as a first target for this analysis.~Globular clusters provide relatively simple stellar populations, so their RR~Lyrae stars share a common distance, age, and bulk chemical composition to a good approximation. This removes several sources of star-to-star variation that would otherwise obscure an intrinsic shift in the PLR zero point. M3 is particularly well suited because it hosts a rich population of RR~Lyrae variables with homogeneous, time-resolved near-infrared photometry. The  survey of Ref.~\cite{2020AJ....160..220B} provides robust mean magnitudes for 233 RR~Lyrae stars and empirical PLRs for the fundamental-mode, first-overtone, and combined samples.

\subsubsection{Stellar modeling}

We implemented the cooling rates in section~\ref{sec:ax-emission} into the stellar structure code MESA (version 23.05.1) \cite{Paxton:2010ji,Paxton:2013pj,Paxton:2015jva,Paxton:2017eie,Paxton:2019lxx,MESA:2022zpy}.~This was used to simulate RR Lyrae stars as follows.

We evolved grids of stars with varying initial mass and wind loss efficiency $\eta$ from the pre-main sequence through the red giant branch, the helium flash, and horizontal branch phases.~We adopted an initial metallicity $Z=7\times10^{-4}$ and initial helium abundance $Y=0.25$, compatible with the old, metal-poor population of M3, which has ${\rm [Fe/H]}\sim -1.5$ \cite{2020AJ....160..220B}.~Wind loss was implemented according to the Reimers prescription \cite{1975MSRSL...8..369R}
\begin{equation}
    \dot M
    =
    -4\times10^{-13}\,\eta\,
    \left(\frac{L}{L_\odot}\right)
    \left(\frac{R}{R_\odot}\right)
    \left(\frac{\msun}{M}\right)
    \,\msun\,{\rm yr}^{-1},
    \label{eq:Reimers}
\end{equation}
where $L$, $R$, and $M$ are the stellar luminosity, radius, and mass.~We varied the initial mass over the range
$0.780 \leq M_{\rm init}/\msun \leq 0.820$, which gives RR~Lyrae models that populate the instability strip at the age of M3, $\tau_{\rm M3}\simeq12.5$~Gyr \cite{2016ApJ...827....2V}, and varied the wind-loss efficiency over the range $0.03\leq\eta\leq0.55$, thereby varying the mass of the hydrogen-rich envelope retained on the horizontal branch and hence the effective temperature, allowing the models to span the instability strip.~The grid spacing was $\Delta M_{\rm init}=5\times10^{-4}\msun$ and $\Delta\eta=0.01$.~We ran separate grids with $\alpha_{26}=\{0,0.1,0.2,0.3,0.5\}$.

During the evolution, radial pulsation periods were computed on the fly using GYRE version 7 \cite{2013MNRAS.435.3406T} via the MESA--GYRE interface.~We retained the fundamental mode, corresponding to RRab variables.

\begin{figure*}[ht!]
    \centering
    \includegraphics[width=0.7\textwidth]{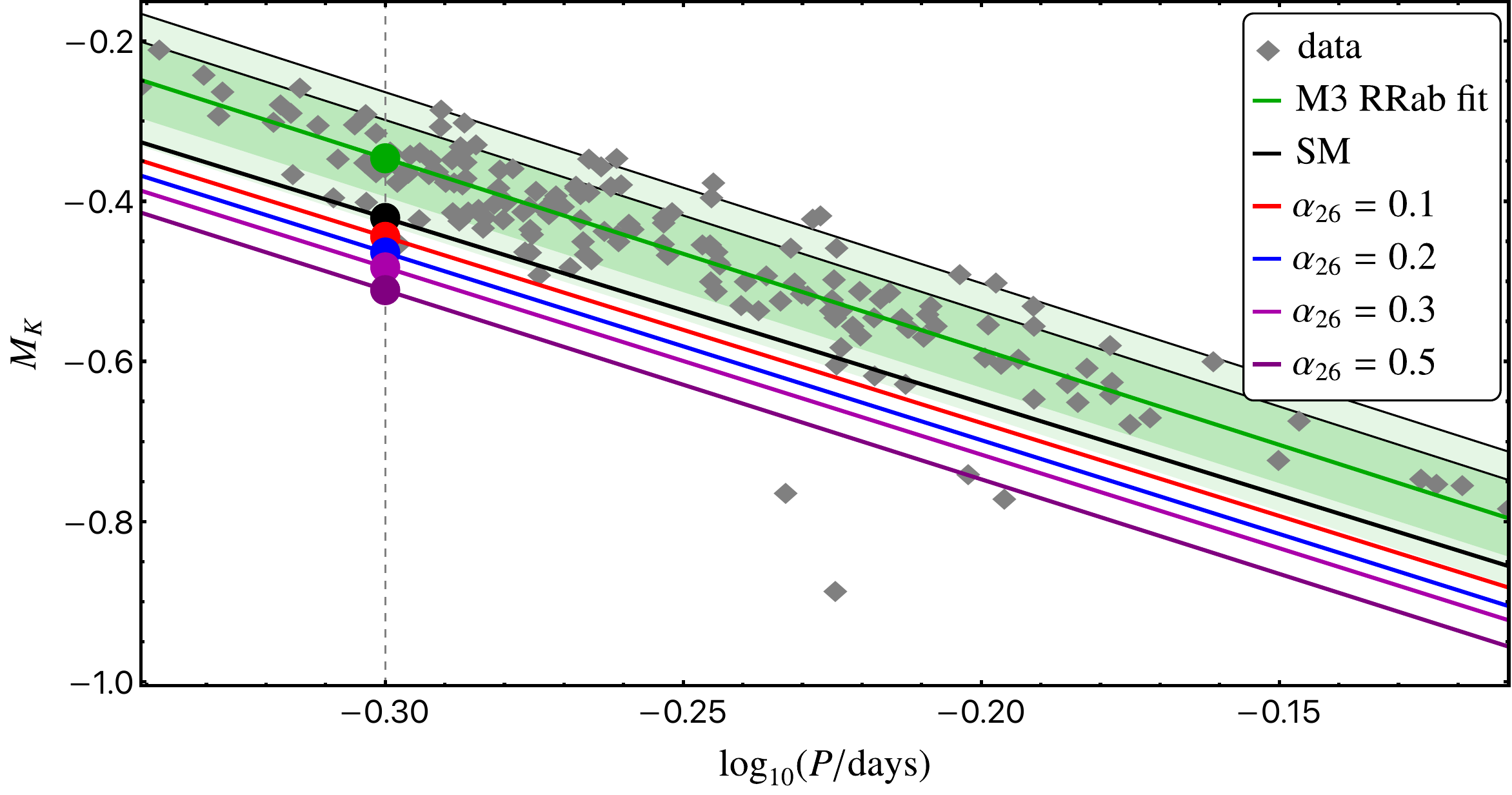}
    \caption{Observed and synthetic RR~Lyrae $K_s$-band period--luminosity relations for M3. Grey diamonds show the clipped RRab sample from Ref.~\cite{2020AJ....160..220B}, converted to absolute magnitudes using 
    {$\mu_{\rm M3}=15.025\pm0.0824$ from the kinematically estimated M3 distance in Ref.~\cite{Baumgardt_2021}.}~The green line shows the best-fit M3 RRab relation. The darker inner band shows the empirical star-to-star scatter, while the lighter outer band shows the $1\sigma$ uncertainty in the absolute zero-point calibration, obtained by combining the formal fit uncertainty with the uncertainty in the kinematic distance modulus in quadrature. The remaining lines show the synthetic fundamental-mode PLRs for the Standard Model and several values of $\alpha_{26}$. Filled circles mark the fitted zero points at the pivot period, $\log P_0=-0.30$, indicated by the vertical dashed line.}
    \label{fig:PL}
\end{figure*}

\subsubsection{Synthetic Period-Luminosity Relation}

For each value of $\alpha_{26}$, we selected RR~Lyrae candidates using the analytic instability-strip boundaries of~\cite{2015ApJ...808...50M}:
\begin{align}
\log T_{\rm blue}
    &= 3.957 - 0.012 \log Z
       - 0.080 \log(L/L_\odot), \\
\log T_{\rm red}
    &= 3.879 - 0.012 \log Z
       - 0.084 \log(L/L_\odot).
\end{align}
These correspond to the first-overtone blue edge and fundamental red edge, respectively, and therefore enclose the full RR~Lyrae instability strip.~We use these as our fiducial analytic boundaries because Ref.~\cite{2015ApJ...808...50M} does not provide an analytic relation for the fundamental-mode blue edge.~All models satisfying $\log T_{\rm red}\leq \log T_{\rm eff}\leq \log T_{\rm blue}$ and lying within $0.1$~Gyr of the adopted M3 age, $\tau_{\rm M3}=12.5$~Gyr were retained.~As a robustness check, we found that restricting the hot edge to temperatures representative of the fundamental-mode blue edge changes the axion-induced zero-point shifts by less than $0.004$ mag.

For each retained model, we converted the equilibrium luminosity to an absolute $K_s$-band magnitude using
\begin{equation}
M_{K_s}=M_{\rm bol}-BC_{K_s}(T_{\rm eff},\log g,{\rm [Fe/H]}),
\end{equation}
with
\begin{equation}
M_{\rm bol}=4.74-2.5\log(L/L_\odot).
\end{equation}
The bolometric corrections were interpolated from the Castelli--Kurucz model-atmosphere tables \cite{2003IAUS..210P.A20C}. We used the table with ${\rm [Fe/H]}=-1.5$ and $\alpha$-enhancement $[\alpha/{\rm Fe}]=+0.2$, appropriate for M3 \cite{2020AJ....160..220B}.~To account for evolutionary residence time, we sample the retained model points with probability proportional to the time interval represented by each point.~As a robustness check on the assumed population weighting, we also reweighted the synthetic samples to reproduce the observed M3 RRab period distribution, finding that the axion-induced zero-point shifts change by less than $0.002$ mag.

We created a synthetic PLR for each $\alpha_{26}$ by fitting the set of retained and processed models to a relation of the form 
\begin{equation}
\label{eq:PLRfit}
M_{K_s}=A+B\left(\log P-\log P_0\right),
\quad \log P_0=-0.30 .
\end{equation}
We adopt $\log P_0=-0.30$ as a convenient reference period for the $K_s$-band PLR.~Using instead the centroid of the observed M3 RRab period distribution changes the axion-induced zero-point shifts by less than $0.003$ mag.

\subsection{Results}
\label{sec:results}

Figure~\ref{fig:PL} shows the synthetic PL relations derived using the pipeline in section~\ref{sec:stellar_response}.~The corresponding fit parameters to Eq.~\eqref{eq:PLRfit} are shown in Table~\ref{tab:plr_fits}.~We also show a fit to the homogeneous M3 $K_s$-band RR~Lyrae catalog of Bhardwaj et al.~\cite{2020AJ....160..220B}, restricting to RRab stars.

\begin{table*}[t]
\centering
\caption{Observed and synthetic RRab/fundamental-mode period--luminosity fits. Each row is fit to Eq.~\eqref{eq:PLRfit} with pivot $\log P_0=-0.30$.~We define $\Delta A\equiv A-A_{\rm SM}$ as the shift relative to the Standard Model prediction. The final column gives the zero-point offset from the M3 RRab fit in units of the absolute zero-point uncertainty, $\sigma_{A,\rm obs}=0.0826$ mag, obtained by combining the formal fit uncertainty with the uncertainty in the kinematic distance modulus from Ref.~\cite{Baumgardt_2021} in quadrature. Negative values correspond to brighter magnitudes.}
\label{tab:plr_fits}
\begin{ruledtabular}
\begin{tabular}{lcccc}
Sample & $A$ & $B$ & $\Delta A$ & offset$/\sigma_{A,\rm obs}$ \\
\hline
M3 RRab & $-0.346$ & $-2.385$ & --       & $0.00$ \\
SM & $-0.421$ & $-2.307$ & $0.000$  & $-0.90$ \\
$\alpha_{26}=0.1$ & $-0.444$ & $-2.323$ & $-0.024$ & $-1.19$ \\
$\alpha_{26}=0.2$ & $-0.464$ & $-2.342$ & $-0.043$ & $-1.43$ \\
$\alpha_{26}=0.3$ & $-0.483$ & $-2.338$ & $-0.062$ & $-1.65$ \\
$\alpha_{26}=0.5$ & $-0.511$ & $-2.362$ & $-0.090$ & $-1.99$ \\
\end{tabular}
\end{ruledtabular}
\end{table*}

The fit was calculated by adopting the distance modulus $\mu_{\rm M3}=15.025\pm0.0824$ from the purely kinematic, RR~Lyrae-independent distance estimate of Ref.~\cite{Baumgardt_2021}, and converting the observed mean magnitudes to absolute magnitudes via $M_{K_s}=K_s-\mu_{\rm M3}$. We retained only stars with well measured mean magnitudes, $\sigma_{K_s}<0.04$ mag, and removed the small number of objects with $K_s<13$, which lie well above the main M3 RR~Lyrae locus and would otherwise dominate the fitted zero point.~We then refit the observed RRab period--luminosity relation directly to Eq.~\eqref{eq:PLRfit}.~The formal uncertainty in the observed M3 RRab fit at the pivot is $0.0053$ mag, while the total absolute zero-point uncertainty is 
{$\sigma_{A,\rm obs}=0.0826$} mag and is therefore dominated by the distance-modulus uncertainty.

As shown in the figure, the predicted SM PLR is consistent with the observed relation at ${0.90}\sigma_{A,\rm obs}$.~The magnitude of this offset, $\simeq {0.074}$ mag, is comparable to theoretical uncertainties encountered in detailed studies of TRGB stellar modeling, e.g.~\cite{Viaux:2013lha,2017A&A...606A..33S,Dennis:2023ldw}.~When ALPs are included, increasing $\alpha_{26}$ results in brighter objects at fixed periods.~For $\alpha_{26}=0.5$, the axion-induced shift relative to the SM is $0.090$ mag, comparable to the absolute zero-point uncertainty $\sigma_{A,\rm obs}= {0.0826}$ mag.~The corresponding total offset from the observed RRab fit is ${1.99}\sigma_{A,\rm obs}$.~Because the synthetic PLRs have somewhat different slopes from the observed M3 RRab relation, we repeated the fits fixing the synthetic slope to the observed value, $B=-2.385$.~The resulting axion-induced zero-point shifts change by less than $0.004$ mag.

\begin{figure}
    \centering
    \includegraphics[width=0.45\textwidth]{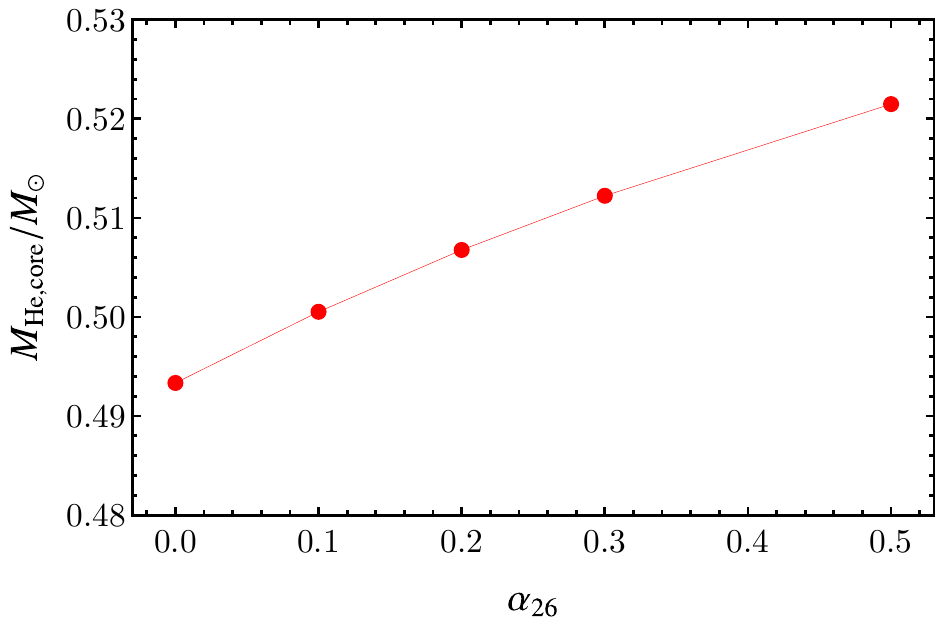}
    \caption{Median helium-core mass of the selected synthetic RR~Lyrae models as a function of $\alpha_{26}$. Axion cooling delays helium ignition on the red giant branch, increasing the helium-core mass of the horizontal-branch models entering the synthetic PLR.}
    \label{fig:HeCoreMass}
\end{figure}

The mechanism underlying the axion-induced brightening is shown in Fig.~\ref{fig:HeCoreMass}. The median helium-core mass of the selected synthetic RR~Lyrae models increases monotonically with $\alpha_{26}$, from $0.493\,M_\odot$ in the Standard Model to $0.521\,M_\odot$ for $\alpha_{26}=0.5$. This $\simeq0.028\,M_\odot$ increase reflects the delayed helium ignition caused by axion cooling on the red giant branch. The larger core mass produces more luminous horizontal-branch models because the HB luminosity is primarily set by the mass of the helium-burning core.

\section{Discussion and Conclusions}
\label{sec:conclusions}

In this work, we have studied the effects of axion-like particles coupled to electrons on the RR~Lyrae period--luminosity relation.~We developed a theoretical pipeline for predicting the PLR using numerical stellar models tailored to M3.~Axion cooling shifts the PLR toward brighter magnitudes, with the shift for $\alpha_{26}=0.5$ comparable to the current absolute zero-point uncertainty, indicating sensitivity to couplings of this order.

The physical mechanism is the same as that underlying TRGB constraints: axion cooling on the red giant branch delays helium ignition and produces a more massive helium core.~In the TRGB, this changes the tip luminosity directly; in RR~Lyrae stars, the larger core mass is instead probed after the helium flash through the brighter horizontal branch and the resulting shift in the near-infrared PLR zero point.

Turning this sensitivity estimate into a robust bound would require addressing two related sources of modeling uncertainty.~First, the inferred sensitivity depends on the absolute calibration of the Standard Model PLR.~Our fiducial Standard Model prediction is shifted toward brighter magnitudes relative to the observations, so modeling choices that change its absolute calibration would also change the inferred sensitivity to axion cooling.~Second, the stellar inputs that affect the helium-core mass and RR~Lyrae PLR may be partially degenerate with the effects of axion cooling and should therefore be marginalized over in a full inference.~In this first study we fixed several modeling choices, including the cluster metallicity and helium abundance, stellar physics parameters and prescriptions, and the bolometric-correction table.~These choices affect either the horizontal-branch luminosity or the conversion from stellar properties to $K_s$-band magnitudes.~The role of such systematics has been studied in detail for TRGB constraints on axion cooling \cite{Viaux:2013lha,2017A&A...606A..33S,Saltas:2022aua,Dennis:2023aam,Dennis:2023ldw}, but an analogous analysis for RR~Lyrae PLRs has not yet been performed.

Several extensions could make the RR~Lyrae analysis more robust.~The study could be repeated for additional globular clusters with homogeneous near-infrared photometry, allowing the dependence on age, metallicity, distance, and HB morphology to be tested directly, while a combined treatment of RRab and RRc stars would make fuller use of the available data.~Given that the present axion--electron sensitivity is weaker than the strongest existing TRGB constraints, obtaining competitive bounds would likely require systems with substantially smaller absolute PLR zero-point uncertainties, particularly in the distance calibration, together with improved control of stellar-model systematics.~The framework may instead be particularly useful for scenarios in which RGB and HB stars respond differently, including other weakly coupled particles and screened modified-gravity theories.~The axion--photon coupling provides one example, since it predominantly affects horizontal-branch evolution rather than the RGB core mass.

\section*{Software}
MESA version~23.05.1, MESASDK version 23.7.3, Mathematica version 12.

\section*{Acknowledgements}

We thank Harry Desmond and Ebraheem Farag for helpful discussions.
~This material is based upon work supported by the National Science Foundation under Grant No.~2207880.~DC is supported by the STFC under Grant No.~ST/T001011/1.~Our simulations were run on the University of Hawai\okina i's high-performance supercomputer KOA.~The technical support and advanced computing resources from University of Hawai\okina i Information Technology Services – Cyberinfrastructure, funded in part by the National Science Foundation MRI award \#1920304, are gratefully acknowledged.~

\section*{Data Availability}

The data and software required to reproduce the results of this work are openly available in the Zenodo repository \href{https://zenodo.org/records/22906030}{https://zenodo.org/records/22906030}, Ref.~\cite{sakstein_2026_22906030}. The repository contains the MESA/GYRE setup, reduced synthetic data, observational and bolometric-correction inputs, and the Mathematica analysis notebook used to reproduce the figures and numerical results.

\bibliography{refs}

\end{document}